\documentclass[nofootinbib,showpacs, prx,twocolumn,preprintnumbers ,amsmath, amssymb, superscriptaddress, aps]{revtex4-2}
\usepackage{lineno}

\usepackage{color}
\usepackage{amsmath,amssymb}
\usepackage[mathscr]{euscript}
\usepackage{pifont}
\usepackage{amssymb}  
\usepackage{bbold}
\usepackage{float}
\usepackage{subfloat}

\usepackage[caption=false]{subfig}
\usepackage{tikz}
\usepackage{makecell}
\usepackage{subfig}
\usepackage{pifont}   
\usepackage{graphicx} 
\usepackage{dcolumn}  
\usepackage{bm}       
\usepackage{multirow} 
\usepackage{placeins}
\usepackage[colorlinks]{hyperref}
\usepackage{mathtools}

\usepackage{dcolumn}
\newcolumntype{d}[1]{D{.}{.}{#1}}
\usepackage{siunitx}

\usepackage{stackengine}

\def\kB{{k_{\rm B}}}
\def\kBT{{k_{\rm B}T}}

\begin{document}

\title{First-Principles Thermodynamics of Al$_{10}$V: An Analytical Treatment of Localized Anharmonic Modes}

\author{Hassan Y. Albuhairan}
\email{halbuhai@andrew.cmu.edu}
\affiliation{Carnegie Mellon University, Department of Physics, Pittsburgh, 15213, PA, USA}
\affiliation{Department of Physics, King Fahd University of Petroleum and Minerals, 31261 Dhahran, Saudi Arabia}

\author{Marek Mihalkovič}
\affiliation{Institute of Physics, Slovak Academy of Sciences, SK-84511 Bratislava, Slovakia}

\author{Michael Widom}
\affiliation{Carnegie Mellon University, Department of Physics, Pittsburgh, 15213, PA, USA}

\date{\today }

\begin{abstract}
Many complex intermetallic structures possess cage-like environments that can host additional guest atoms. In Al$_{10}$V, these atoms give rise to low-frequency, localized vibrations (Einstein modes) that dominate the thermodynamic response at low temperature. They become imaginary under volume expansion as temperature rises, invalidating the harmonic approximation. We develop a framework to incorporate these strongly anharmonic vibrational modes into first-principles thermodynamic calculations. By explicitly modeling the cage potential and solving the associated Schrödinger equation numerically, we compute the full anharmonic free energy contribution and demonstrate its impact on thermodynamic phase stability. Our results reproduce key experimental signatures, including the anomalous rise in the thermal expansion coefficient and specific heat at low temperatures, and reveal that the presence of guest atoms is essential to stabilizing the Al$_{10}$V phase at elevated temperatures. 
\end{abstract}

\maketitle 

\section{Introduction}

Clathrates~\cite{clathrates,Goto2004}, fullerenes~\cite{Curl1991,fullerenes}, skutterudites~\cite{Sales1997,Keppens1998}, and other complex intermetallics~\cite{Borides,Perovskite}—including quasicrystals~\cite{Tamura_2002} and metal–organic frameworks~\cite{MOF}—commonly feature cage-like structures that host one or more guest atoms. The dynamics of the guest atoms, and their interactions with the cage, often yield interesting physical properties such as phonon lifetimes~\cite{BaGeAu}, thermoelectric properties~\cite{Rogl}, heat capacity anomalies~\cite{AV2Al20}, and superconductivity~\cite{MV2Al20}.

\begin{figure}[b!]
  \centering
\stackinset{l}{0.9cm}{t}{0.3cm}{{\Large (a)}}{\includegraphics[trim = 1.2cm 0cm 1.2cm 0cm, clip, width=.45\textwidth]{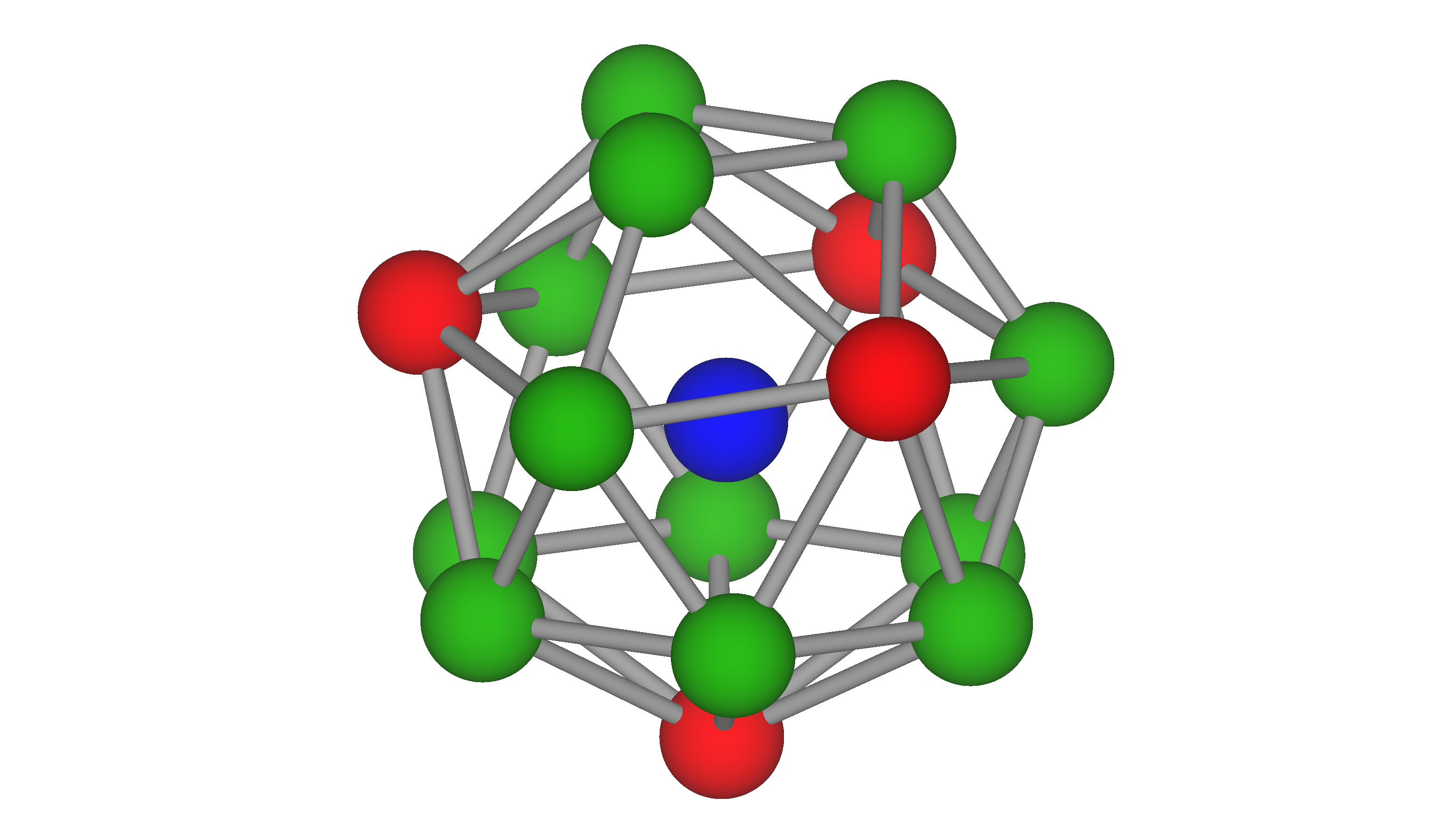}}
\includegraphics[width=0.45\textwidth,clip]{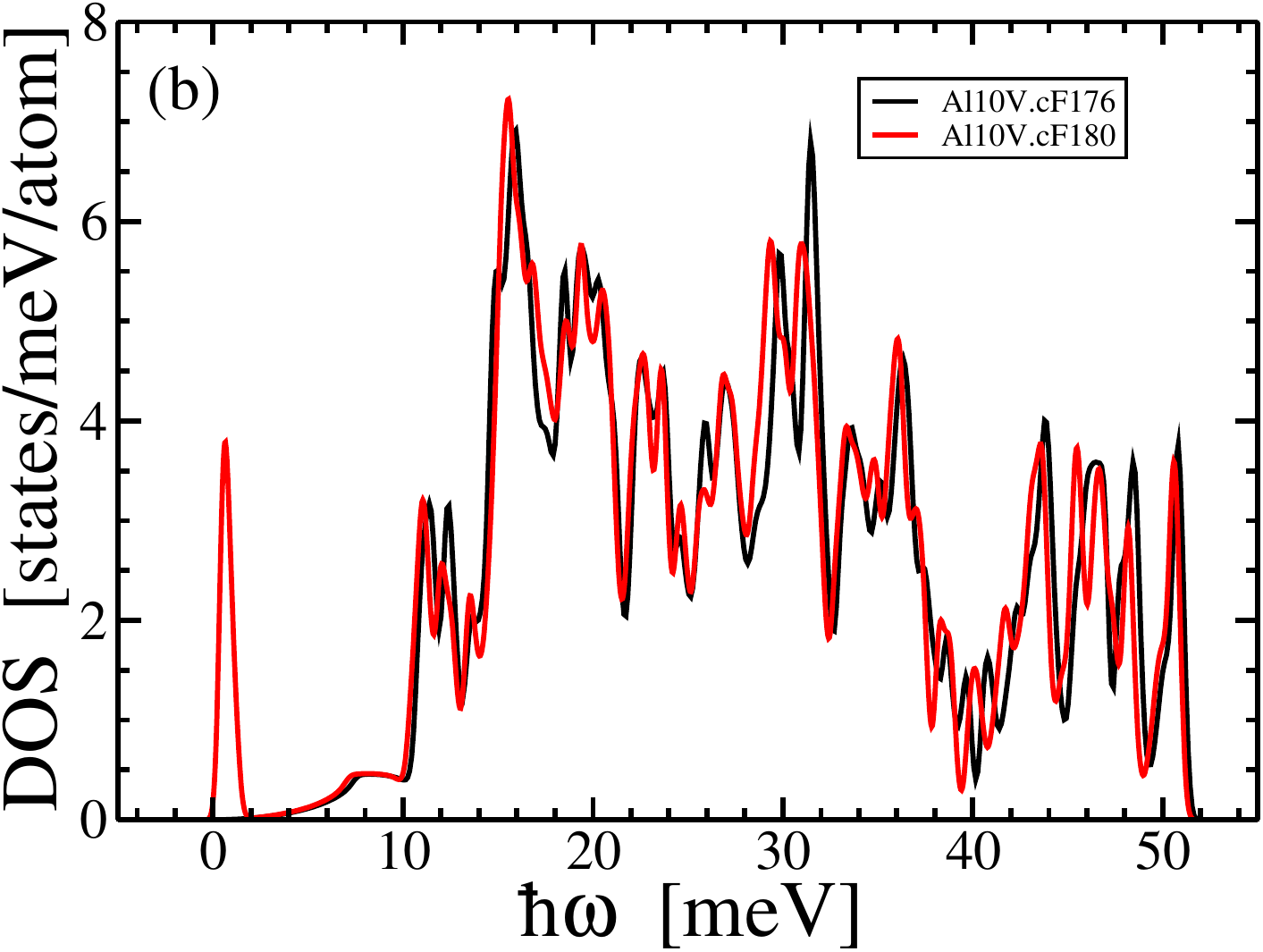}
\caption{(a) Cage structure of Al$_{10}$V: green spheres indicate Al atoms at the $96g$ Wyckoff site, red spheres represent Al atoms at the $16c$ site forming a tetrahedron, and the void is occupied by an Al atom at the $8a$ site (blue sphere). (b) Vibrational density of states for Al$_{10}$V.cF176 (black) and Al$_{10}$V.cF180 (red), computed using \textsc{Phonopy}~\cite{phonopy1,phonopy2}.}
\label{fig:structure}
\end{figure}

The focus of our present study, Al$_{10}$V~\cite{Brown1957,Ray1957}, is such an example. As first reported by Caplin and co-workers in the 1970s~\cite{Caplin1973,Caplin1978} Al$_{10}$V exhibits anomalous low-temperature behavior with the specific heat and electrical resistivity following an Einstein-like temperature dependence~\cite{Einstein1907}. This was attributed to the vibrations of loosely bound Al atoms occupying the cage environments, where low-frequency local modes become thermally activated and dominate the vibrational spectrum over the relevant temperature range.

The unit cell of Al$_{10}$V is face-centered cubic and contains 160 Al atoms and 16 V atoms. It includes eight cages per unit cell (two per primitive cell); each cage, illustrated in Fig.~\ref{fig:structure}(a), is a 16-atom Friauf polyhedron~\cite{Friauf} composed of 12 Al atoms at the $96g$ Wyckoff site (green spheres) and 4 Al atoms at the $16c$ site (red), which form a tetrahedron. The V atoms act as ``glue'' to hold the cages within a diamond-like structure. The cage void at site $8a$ may be occupied by an Al atom (blue)—referred to in this work as the guest atom—although the precise occupancy remains experimentally uncertain. The structure and bonding is described in detail in Ref.~\cite{Jahntek2003}. In this study, we focus on a structure in which half of the cages are filled, denoted Al$_{10}$V.cF180 (using the Pearson symbol), and we compare it with the empty-cage structure, Al$_{10}$V.cF176, to assess the effect of the guest atoms.

Figure~\ref{fig:structure}(b) shows the vibrational density of states (vDOS) for Al$_{10}$V.cF176 (black curve) and Al$_{10}$V.cF180 (red curve). While the overall features are similar, the Al$_{10}$V.cF180 spectrum displays a pronounced low-frequency peak that we identify as a local mode arising from the guest atom. We identify the Einstein frequency $\omega_{\rm E}\approx 0.65~\rm{meV}/\hbar$ and Einstein temperature $\Theta_{\rm E}=\hbar\omega_{\rm E}/k_{\rm B}\approx 7.5$K. This mode is responsible for the unusual thermal behavior observed in Al$_{10}$V. Due to its low frequency—and the fact that it becomes imaginary upon volume expansion—the mode is strongly anharmonic.  When a larger guest atom occupies the $8a$ site\footnote{The international crystallographic structure database lists 130 compounds belonging to the Mg$_3$Cr$_2$Al$_{18}$ prototype, all isostructural with Al$_{10}$V. Among these, 15 compounds are derived from Al$_{10}$V with various guest atoms occupying the void site, including several lanthanide and actinide rare earth elements.}, such as Gd in GdV$_2$Al$_{20}$ \cite{Verbovytsky2007} or Eu in EuV$_2$Al$_{20}$ \cite{Chi2008}, the guest atom mode shifts upwards, indicating that the extreme low frequency in Al$_{10}$V must be due to exceptionally weak interaction of the Al guest atom with the cage.

We present a general analytical approach for treating strongly anharmonic vibrational modes when their atomic origin is known, as is the case for Al$_{10}$V, where the anharmonicity arises from the dynamics of guest atoms. We apply this method to predict thermal properties of Al$_{10}$V from first principles.

This paper is organized as follows. We begin by outlining the computational settings and the Al–V phases considered. We then present the theoretical framework, including the free energy formulation, our treatment of the anharmonic mode, the construction of the cage potential, the numerical solution of the associated Schrödinger equation, and the classical treatment at high temperatures. The results section follows, where we examine the thermal expansion and specific heat at low temperatures, and then present the computed Al–V phase diagram over the range $0 < T < 1000$ K. We conclude with a summary of our main findings.

\section{Calculation details}

We performed first-principles total‐energy calculations using the Vienna Ab initio Simulation Package (\textsc{Vasp}~\cite{vasp1,vasp2,KJ_PAW}) within the PW91~\cite{Perdew1993} generalized‐gradient approximation. The PBE~\cite{Perdew1996} functional yields different results and fails to reproduce the experimentally observed anomalous behavior of Al$_{10}$V. We discuss the use of these functionals, in addition to the HSE06~\cite{hse06} hybrid functional, in Appendix \ref{app:dft_functionals}. Our Al pseudopotential is the standard one with valence 3; our vanadium potential has valence 13, as it treats the 3$s$ and 3$p$ semicore levels as valence. All atomic positions and lattice parameters were fully relaxed, using a fixed plane‐wave cutoff energy of 340 eV (compared with the default values of 240.437 eV for Al and 263.695 eV for V) in order to minimize the Pulay stress, and with the projection operators evaluated in reciprocal space to achieve accurate forces. We used accurate precision settings to avoid wrap-around errors. Phonon properties were then obtained via density‐functional perturbation theory (DFPT). 

For the Al--V phase diagram calculations, we include all known Al--V phases that may compete with Al$_{10}$V. The particularly stable Al$_3$V phase divides the Al--V compositional phase space, so we restrict our thermodynamic analysis to phases with vanadium content at most that of Al$_3$V. Table \ref{tab:phases} summarizes the considered phases, along with their number of atoms per cell and $k$‐point densities at which energies were sufficiently converged. \textsc{VASP} input and output files are provided in the supplemental materials~\cite{SuppMat}.

\begin{table}[H]
  \centering
  \caption{Crystal structures, number of atoms per cell, and $k$-point meshes used for Brillouin zone sampling in this study.}
  \label{tab:phases}
  \begin{tabular}{l c c}
    \hline
    \textbf{Structure} & \textbf{Atoms/cell} & \textbf{$k$-point mesh} \\
    \hline
    Al.cF4                    & 32  & 5×5×5 \\
    Al$_{10}$V.cF176          & 44  & 5×5×5 \\
    Al$_{10}$V.cF180          & 45  & 5×5×5 \\
    Al$_{12}$V.cI26           & 26  & 6×6×6 \\
    Al$_{23}$V$_4$.hP54       & 54  & 6×6×2 \\
    Al$_3$V.tI8               & 64  & 6×6×3 \\
    Al$_{45}$V$_7$.mC104      & 52  & 6×6×5 \\
    V.cI2                     & 54  & 5×5×5 \\
    \hline
  \end{tabular}
\end{table}

\section{Theory}

\subsection{Free energy calculations}

To obtain temperature-dependent thermodynamic properties, we evaluate the Helmholtz free energy of each phase as a sum of electronic, vibrational, and (where relevant) configurational contributions. The Helmholtz free energy is given by~\cite{WidomJMR2018}
\begin{equation}\label{eq:Helmholtz-free-energy}
    F(T,V) = E(V) + F_\mathrm{vib}(T,V) + F_\mathrm{conf}(T) + F_\mathrm{elect}(T),
\end{equation}
where all quantities are given per atom:
\begin{itemize}
  \item \( E(V) \) is the DFT total electronic energy of the fully relaxed structure at volume \( V \);
  
  \item \( F_\mathrm{vib}(T,V) \) is the harmonic vibrational free energy~\cite{Fultz2010},
  \begin{equation}\label{eq:harmonic-free-energy}
    F_\mathrm{vib}(T,V)
    = \kBT \sum_{i=1}^{3N-3} 
      \ln\left[2\sinh\left(\frac{\hbar \omega_i(V)}{2\kBT}\right)\right],
  \end{equation}
  where \( \omega_i(V) \) is the $i$-th phonon frequency at volume \( V \), and \( N \) is the number of atoms in the cell. The summation excludes the three zero-frequency translational modes;

  \item \( F_\mathrm{conf}(T) = -T S_\mathrm{conf}(T) \) is the configurational free energy. For partial occupation over \( N \) equivalent sites with occupation fraction \( p \), the configurational entropy is
  \begin{equation}
    S_\mathrm{conf} = -N \kB \left[p \ln p + (1 - p) \ln(1 - p)\right].
  \end{equation}
  In the case of Al$_{10}$V.cF180, with half-filled voids, this reduces to
  \begin{equation}
    \frac{S_\mathrm{conf}}{\text{atom}} = \frac{2}{45} \kB \ln 2;
  \end{equation}
  \item The electronic free energy $F_{\rm elect}$ is obtained from integrals over the electronic density of states~\cite{Felect}.
\end{itemize}

Following the quasiharmonic approach, we determine the zero-pressure equilibrium Gibbs free energy \( G(T) \) at each temperature by performing a series of constant-volume calculations and minimizing \( F(T,V) \) with respect to volume:
\begin{equation}
  G(T) = \min_V F(T, V).
\end{equation}
In Sec.~\ref{subsection:low-T-response}, we present the low-temperature thermal response. We begin with the linear thermal expansion coefficient, \( \alpha(T) \), which we evaluate using
\begin{equation}\label{eq:alpha}
    \alpha(T) = \frac{1}{a_\mathrm{eq}(T)} \frac{d a_\mathrm{eq}(T)}{dT},
\end{equation}
where \( a_\mathrm{eq}(T) \) is the lattice constant that minimizes the Helmholtz free energy at temperature \( T \), {\em i.e.},
\begin{equation}
a_\mathrm{eq}(T) = \operatorname*{arg\,min}_a F(T, V(a)).
\end{equation}
We also compute the specific heat at constant pressure using
\begin{equation} \label{eq:cp}
    C_p(T) = -T \frac{\partial^2 G(T)}{\partial T^2}.
\end{equation}
In Sec.~\ref{subsection:phase_diagram}, we present the binary Al--V phase diagram. The formation free energy of a compound Al$_m$V$_n$ is computed as
\begin{equation}\label{eq:formation-free-energy}
    \Delta G(T,x) = G^{\mathrm{Al}_m \mathrm{V}_n}(T)\\
    - \bigl[(1 - x) G^{\mathrm{Al}}(T)
    + x G^{\mathrm{V}}(T)\bigr],
\end{equation}
where \( x = n / (m + n) \) is the vanadium atomic fraction. Phase stability is then determined by computing the convex hull of \( \Delta G(T,x) \) values for all competing phases at each temperature $T$.

\subsection{Anharmonic mode}

\begin{figure}[b!]
\centering
\includegraphics[width=\linewidth,clip]{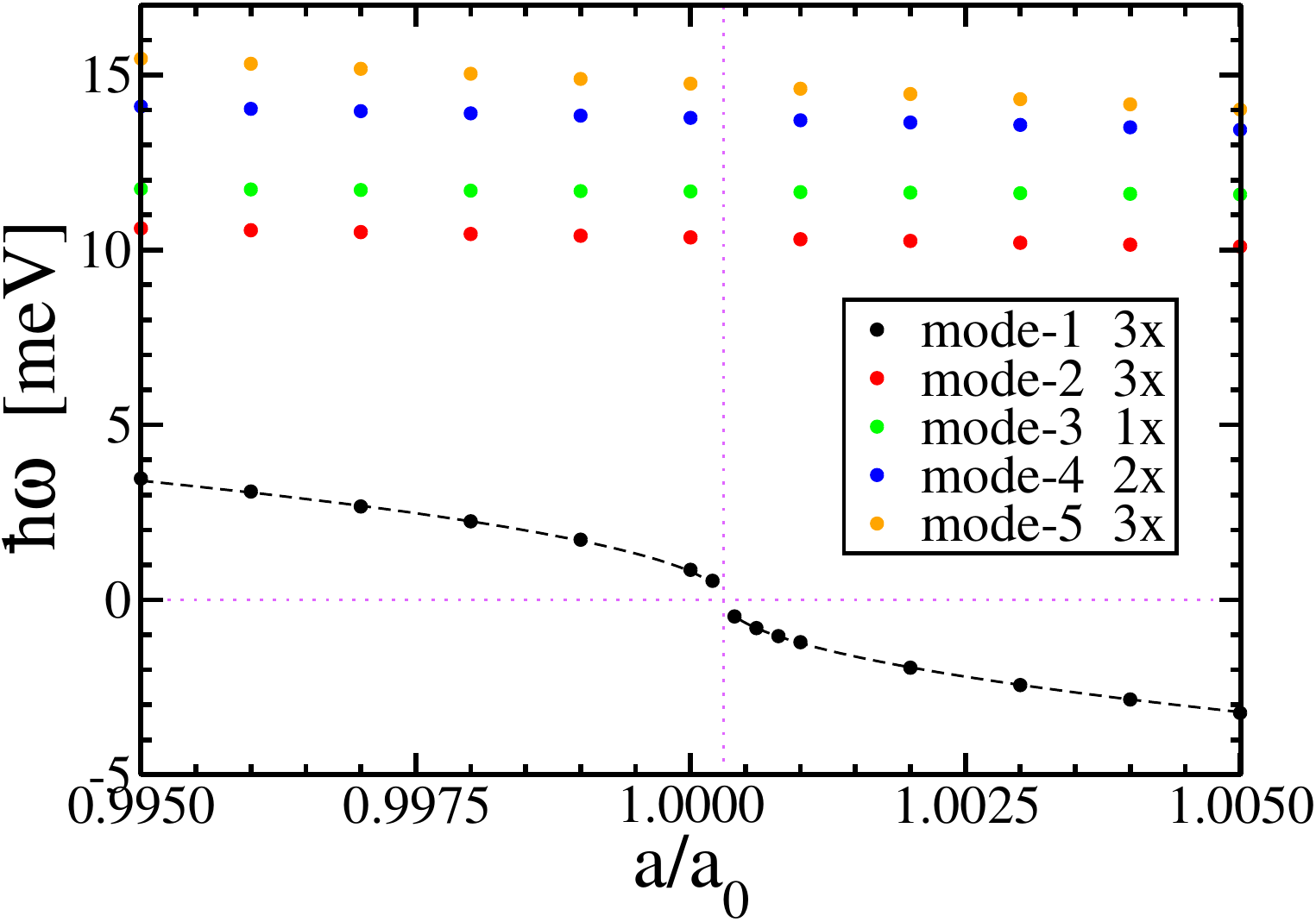}
\caption{Evolution of the vibrational mode frequencies at the $\Gamma$-point as a function of cell volume, expressed in terms of the lattice constant scaling $a/a_0$. Negative values on the $y$-axis indicate imaginary frequencies. The black dashed line represents a fit with a square-root singularity. Dotted lines locate the zero crossing at $a/a_0=1.0003$.}
\label{fig:FreqVsVol}
\end{figure}

\begin{figure*}[t!]
\centering
\stackinset{l}{3cm}{t}{1.5cm}{\includegraphics[width=.15\textwidth]{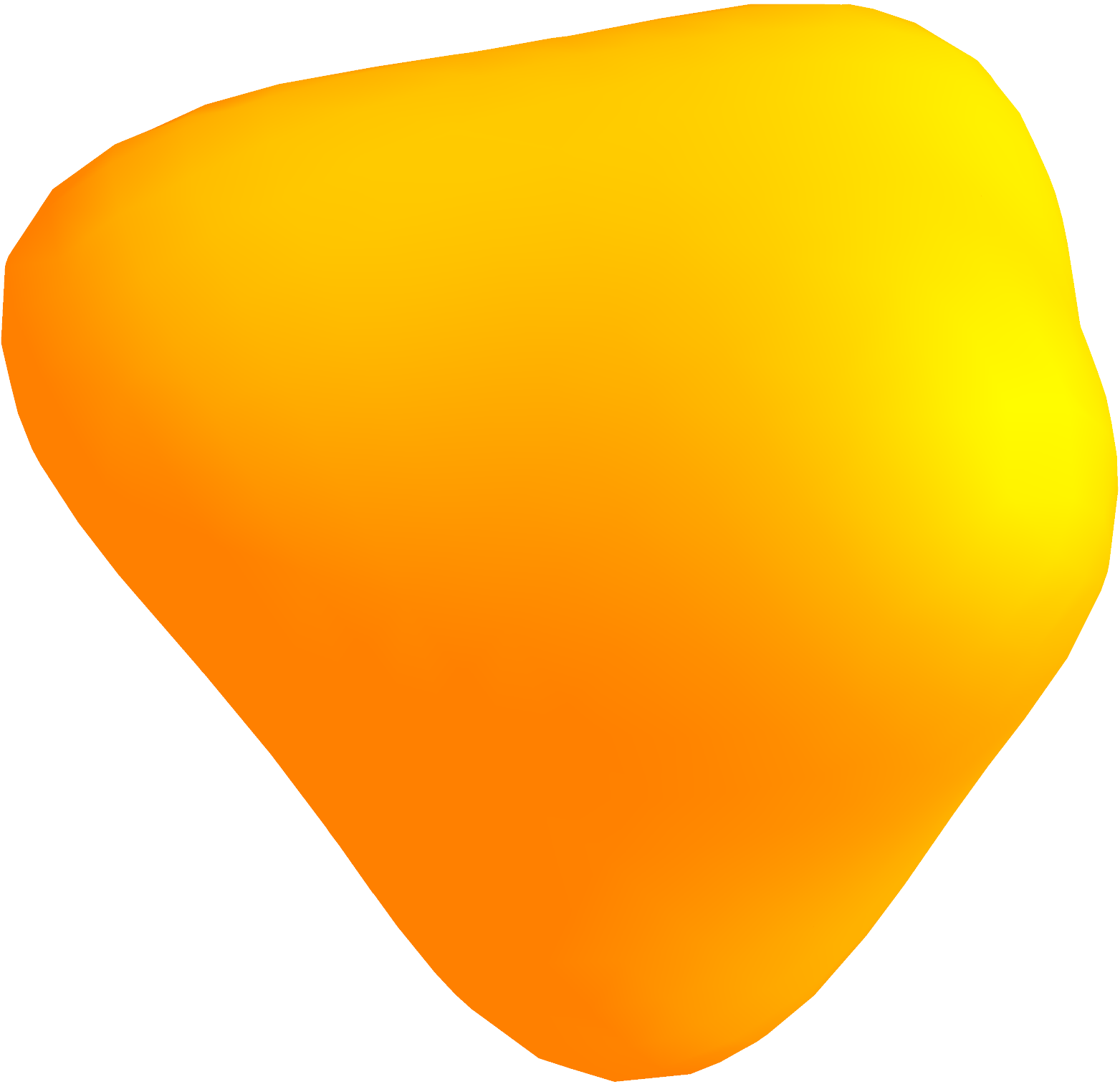}}{\includegraphics[width=.45\textwidth]{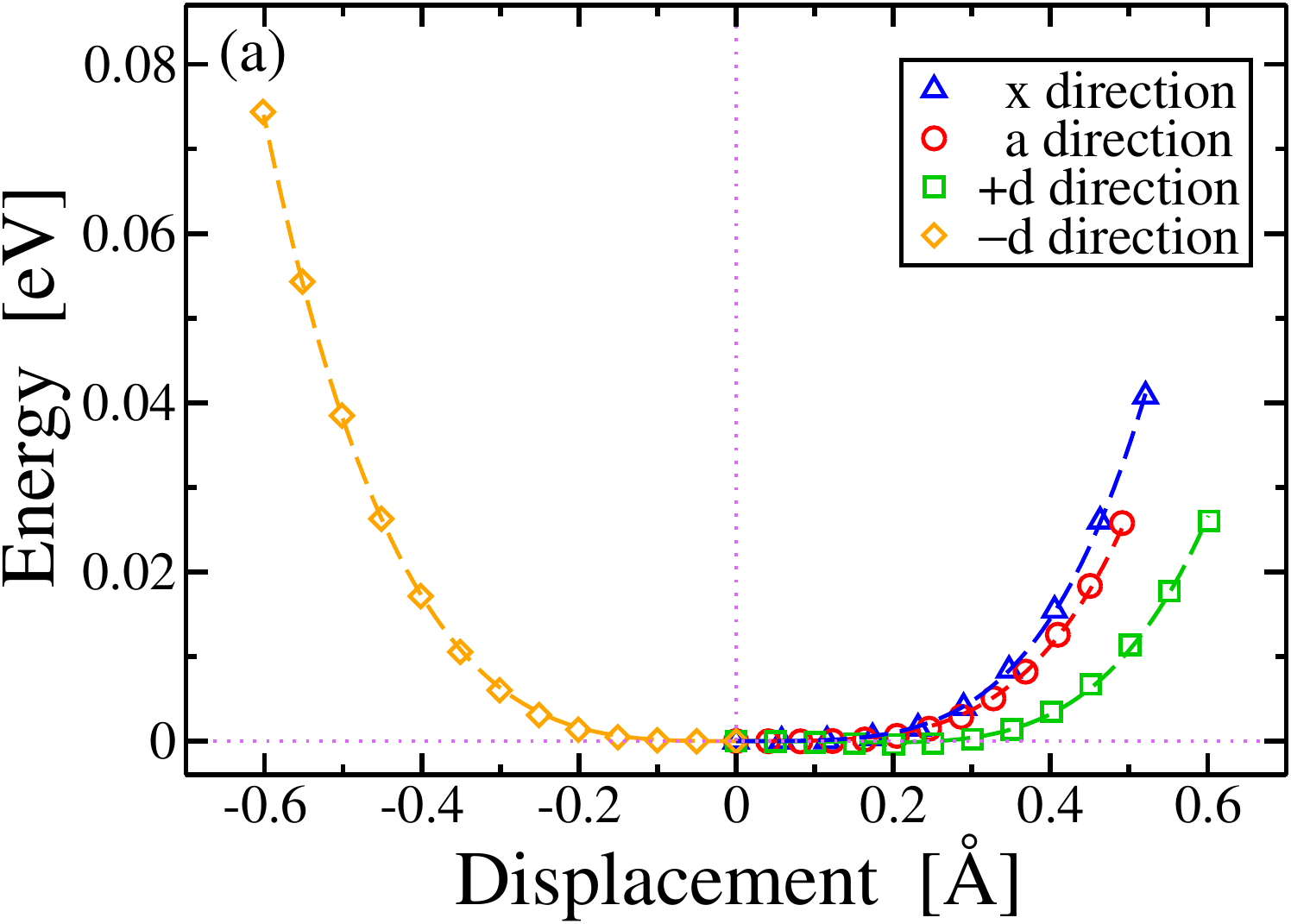}}
\hspace{0.5cm}
\stackinset{l}{3cm}{t}{1.1cm}{\includegraphics[width=.15\textwidth]{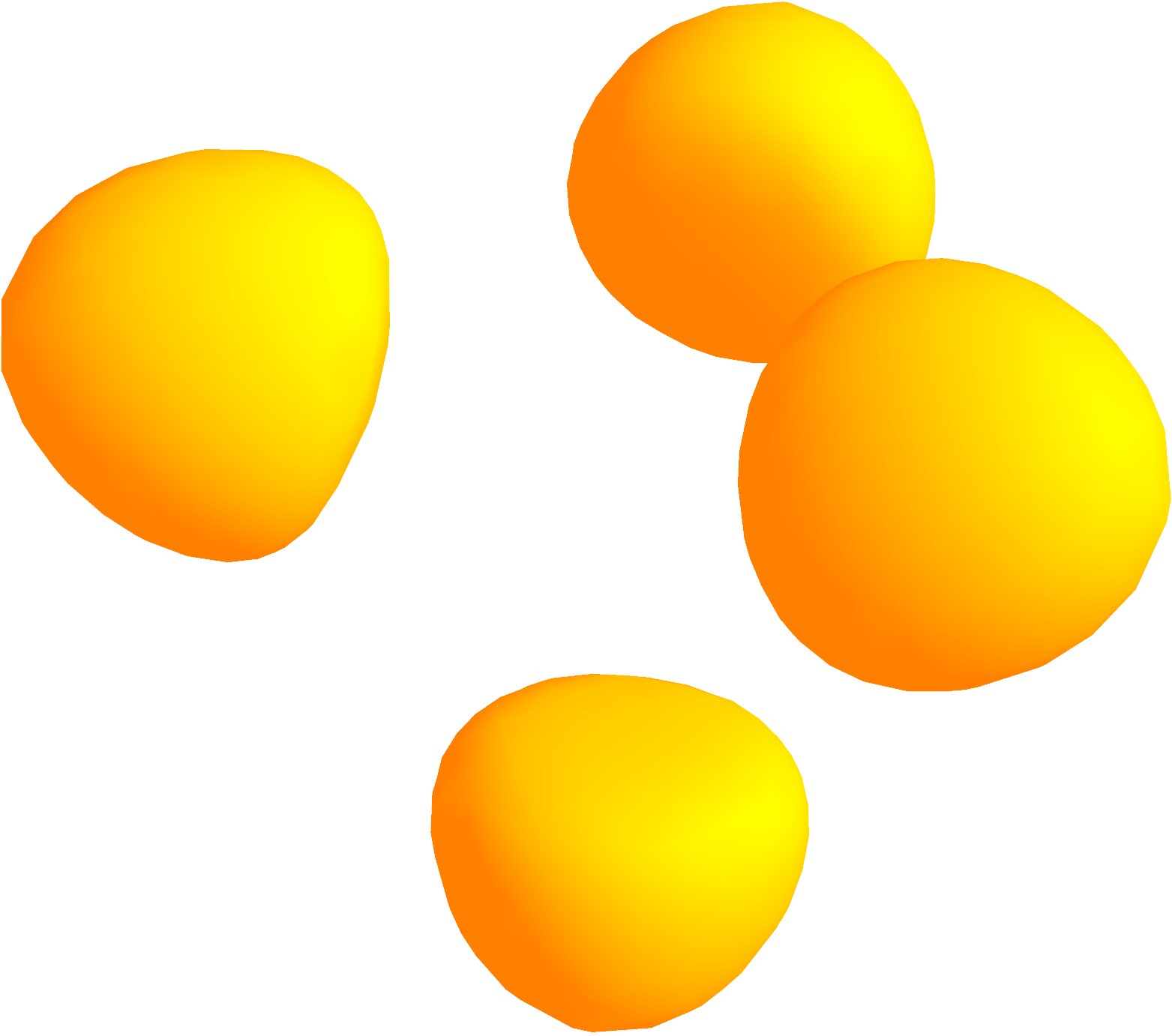}}{\includegraphics[width=.45\textwidth]{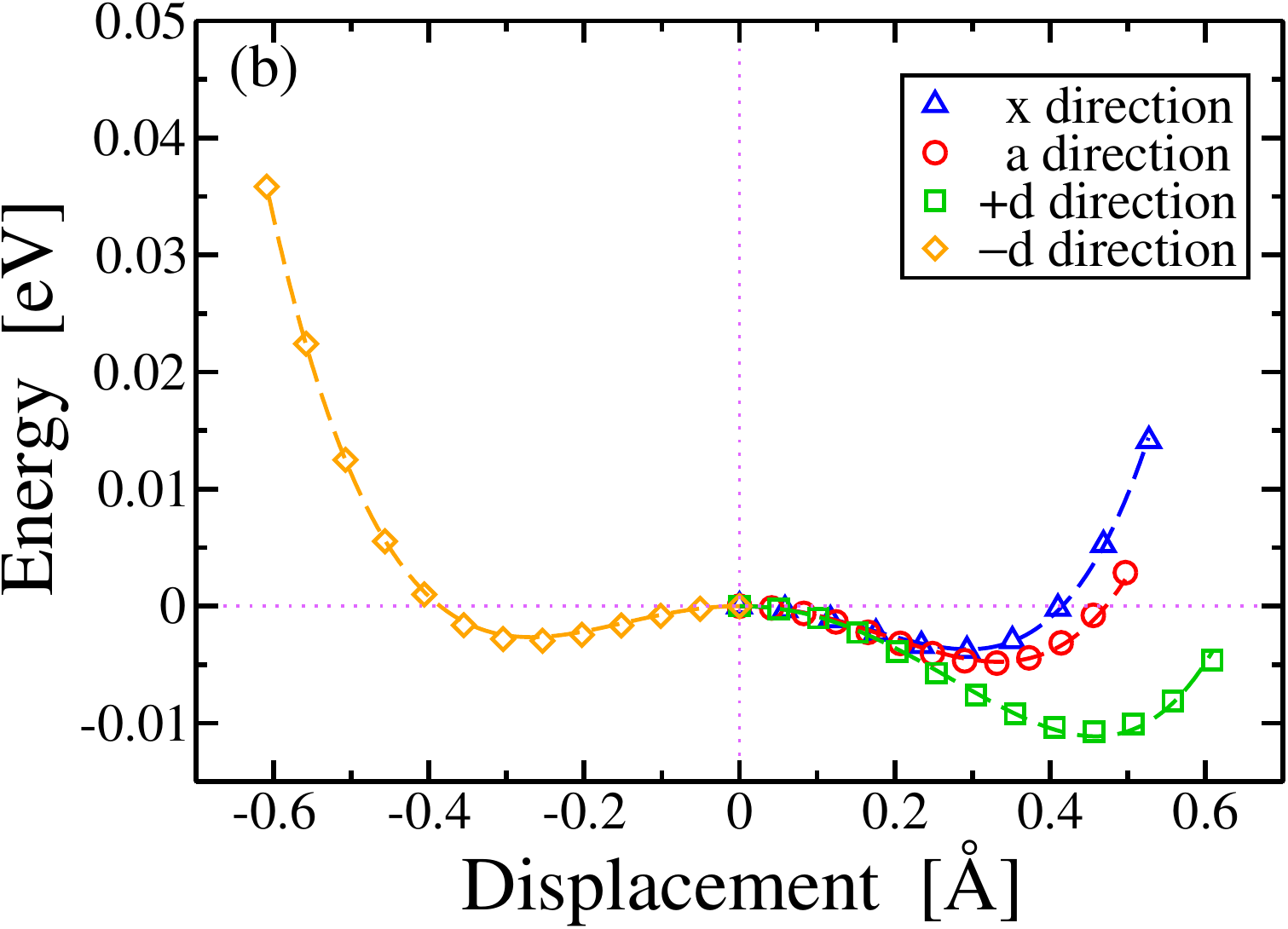}}
\caption{Energy as a function of guest atom displacement along various directions for (a) the relaxed volume with $a/a_0 = 1.000$ and (b) an expanded volume with $a/a_0 = 1.012$. Insets show potential isosurfaces of $U(x,y,z)=U_{\mathrm{min}}+\delta$, where $\delta=0.005$ eV, for the same respective volumes.}
\label{fig:Potential}
\end{figure*}

Figure~\ref{fig:FreqVsVol} shows the lowest non-zero phonon frequencies at the phonon $\Gamma$ point versus the cell volume, expressed as the lattice-constant scaling $a/a_0$. Here $a_0=14.4762$~\AA~ is the relaxed lattice constant ($a_\text{exp} \approx14.5$~\AA~\cite{Brown1957,Ray1957}). The first non-zero mode, corresponding to the local vibration of the guest atom, is plotted as black dots.  It is 3-fold degenerate. As the volume increases, its frequency softens and becomes imaginary (represented by negative values) at $a=a^* \approx 1.0003\,a_0$.  The dashed line is the fit $\omega(a) = \operatorname{sgn}(a^* - a_0)\,\sqrt{k\,|a^* - a|}$, with $k = 11.3$ determined from the data. Because this local mode goes imaginary, Eq.~\eqref{eq:harmonic-free-energy} no longer applies. We therefore separate the guest contribution and write
\begin{equation}\label{eq:corrected-harmonic-free-energy}
  F_\mathrm{vib}(T,V)
    = F_\mathrm{guest}(T,V)
    + \kBT \sum_{i=4}^{3N-3} 
      \ln\bigl[2\sinh\bigl(\tfrac{\hbar\omega_i(V)}{2\kBT}\bigr)\bigr]\,,
\end{equation}
where
\begin{equation}\label{eq:void-atom-free-energy}
  F_\mathrm{guest}(T,V)
  = -\kBT \ln\biggl[\sum_j e^{-\epsilon_j/\kBT}\biggr]
\end{equation}
is the quantum free energy of the guest atom in its cage and \( \epsilon_j(V) \) are the eigenvalues of the Schrödinger equation. Equation~\eqref{eq:void-atom-free-energy} is valid in the limit of an infinite number of eigenstates. For any finite set, however, the truncated sum systematically underestimates the free energy at high temperatures, where the neglected high-energy states become thermally accessible. We illustrate the impact of this truncation on \( F_\mathrm{guest}(T, V) \) in Appendix~\ref{app:quantum_vs_calssical_freeEnergy}. 

We solve the Schrödinger equation numerically using \textsc{Mathematica}, imposing Dirichlet boundary conditions such that the wavefunction vanishes at the boundaries of a cubic region with side length $2r=3$ Å—that is, 1.5 Å from the center of the potential in each direction. We compute the lowest $n=1000$ eigenvalues. The code used to perform these calculations is provided in the supplemental materials~\cite{SuppMat}. The quantum free energy of the guest atom, $F_Q= F_\mathrm{guest}(T,V)$ defined by Eq.~\eqref{eq:void-atom-free-energy}, is employed for \( T < 250\,\mathrm{K} \). At higher temperatures more than $n=1000$ eigenvalues are required to converge the free energy, and instead for \( T \geq 250\,\mathrm{K} \) we use the classical approximation:
\begin{equation}\label{eq:classical-free-energy}
  F_C(T,V)
  = -\kBT \left[
    \tfrac{3}{2} \ln\left(\tfrac{m \kBT}{2\pi \hbar^2}\right)
    + \ln Z_c(T,V)
  \right],
\end{equation}
where
\begin{equation}\label{eq:classical-partition-function}
 Z_c(T,V) = \int_{-r}^{r} \int_{-r}^{r} \int_{-r}^{r}
e^{-U(x,y,z)/\kBT} dx\,dy\,dz
\end{equation}
is the positional partition function. Here, \( U(x,y,z) \) is the volume-dependent cage potential energy and the integration is performed over a cube of side length \( 2r \). While the classical calculation does not involve boundary conditions, we use the same value of \( r \) as in the quantum calculation—where it defines the Dirichlet boundary—to ensure consistency in the spatial domain over which the potential is sampled. We assess the validity of the quantum and classical free energy expressions for the guest atom in Appendix~\ref{app:quantum_vs_calssical_freeEnergy}.

\subsection{Cage potential}

To probe the potential energy landscape of the cage experienced by the guest atom, we computed DFT energies as the guest atom was displaced along several high-symmetry directions. All other Al atoms are relaxed, holding lattice parameters and V atoms fixed.

Figure~\ref{fig:Potential}~(a) shows the total DFT energy (zeroed at the cage center) versus the guest atom displacement along the Cartesian $x$-, primitive lattice vector $a$- ($\propto x+y$), body diagonal $+d$- ($x+y+z$), and $-d$-directions.  The nearly flat curves near the origin reveal a strongly anharmonic potential. The softer rise along $+d$ towards one of the tetrahedral vertex atoms shown in Fig.~\ref{fig:structure}~(a) versus $-d$ (toward a tetrahedral face) confirms pronounced anisotropy. Recall the tetrahedral atoms lie closer to the void center than do the rest of the cage, so the softest displacement is towards the closest cage atom.

Figure~\ref{fig:Potential}~(b) shows the energy profile at an expanded volume (\( a/a_0 = 1.012 \)), where the potential becomes non-convex at the origin consistent with the emergence of imaginary phonon modes at the \(\Gamma\)-point, as seen in Fig.~\ref{fig:FreqVsVol}.

We fit the computed potential energy data to a basis of tetrahedral (cubic) invariants~\cite{Bethe1947,WORMER2001,Kareem} up to fourth order. The fitted potential takes the form
\begin{multline}\label{eq:potential-fit}
U(x,y,z)
  = a_2\,(x^2 + y^2 + z^2)
  + a_4^{\mathrm{iso}}\,(x^2 + y^2 + z^2)^2 \\
  + a_4^{\mathrm{aniso}}\bigl[x^4 + y^4 + z^4
    - 3(x^2 y^2 + x^2 z^2 + y^2 z^2)\bigr]
  + a_3\, x y z,
\end{multline}
where \( a_2 \), \( a_4^{\mathrm{iso}} \), \( a_4^{\mathrm{aniso}} \), and \( a_3 \) are the coefficients of the quadratic, isotropic quartic, anisotropic quartic, and cubic (inversion-breaking) terms, respectively. For comparison, Ref.~\cite{Safarik2012} modeled the cage potential using a sixth-order interatomic form, with negligible contributions from both harmonic and quartic terms. Ref.~\cite{Koza2015}, in contrast, used a simpler isotropic model with only quadratic and quartic terms.

The coefficients in Eq.~\eqref{eq:potential-fit} are obtained from a global fit to energy data along all displacement directions. The dashed lines in Fig.~\ref{fig:Potential} show directional cuts through the fitted potential \( U(x,y,z) \), revealing that the quadratic term is weakly positive at the relaxed volume and strongly negative at the expanded volume. The full set of fitted coefficients is listed in Table~\ref{tab:coeffs}. 

\begin{table}[H]
\centering
\begin{tabular}{l|rrrr}
$a/a_0$ & $a_2$ & $a_4^{\mathrm{iso}}$ & $a_4^{\mathrm{aniso}}$ & $a_3$ \\
\hline
1      &  0.0035 & 0.4434 & 0.1011 & -0.5673 \\
1.012  & -0.0854 & 0.4054 & 0.0886 & -0.4571 \\
\hline
\end{tabular}
\caption{Coefficients of the cage potential fit to the functional form $U(x,y,z)$ of Eq.~(\ref{eq:potential-fit}). Units are eV and~\AA.}
\label{tab:coeffs}
\end{table}

Analysis of the fitted potential surface reveals four off-center minima located at the Wyckoff positions \( 32e \), which lie along the directions connecting the guest atom to the four tetrahedral Al atoms, illustrated in Fig.~\ref{fig:structure}. These points collectively form a smaller tetrahedron within the cage. The \( 32e \) sites have the form \( (x, x, x) \) with \( x > 0 \). For the relaxed volume, the minima occur at \( x = 0.112 \)~\AA{} with a well depth of 0.132~meV, whereas at the expanded volume \( a/a_0 = 1.012 \), the minima shift to \( x = 0.265 \)~\AA{} and the well depth increases to 11.1~meV. The insets in Figs.~\ref{fig:Potential}(a) and \ref{fig:Potential}(b) show the potential isosurfaces near these minima, defined by \( U(x, y, z) = U_{\mathrm{min}} + \delta \), with \( \delta = 0.005 \)~eV corresponding to temperature $T=\delta/k_{\rm B}=58$K.


\section{Results and discussion} \label{results}

\subsection{Low-temperature thermal response}\label{subsection:low-T-response}

\begin{figure*}[t!]
\centering
\includegraphics[width=.45\textwidth,clip]{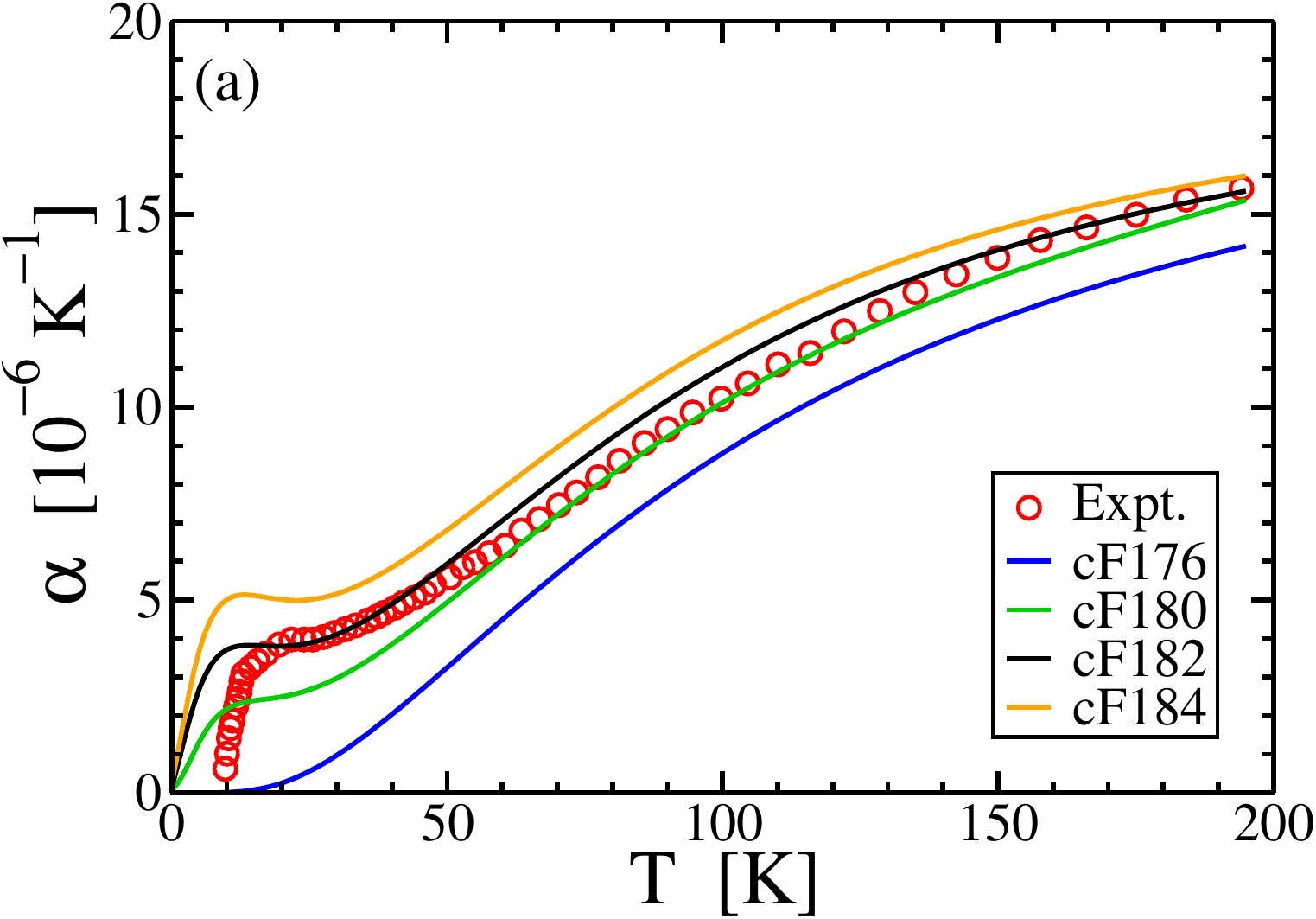}
\includegraphics[width=.45\textwidth,clip]{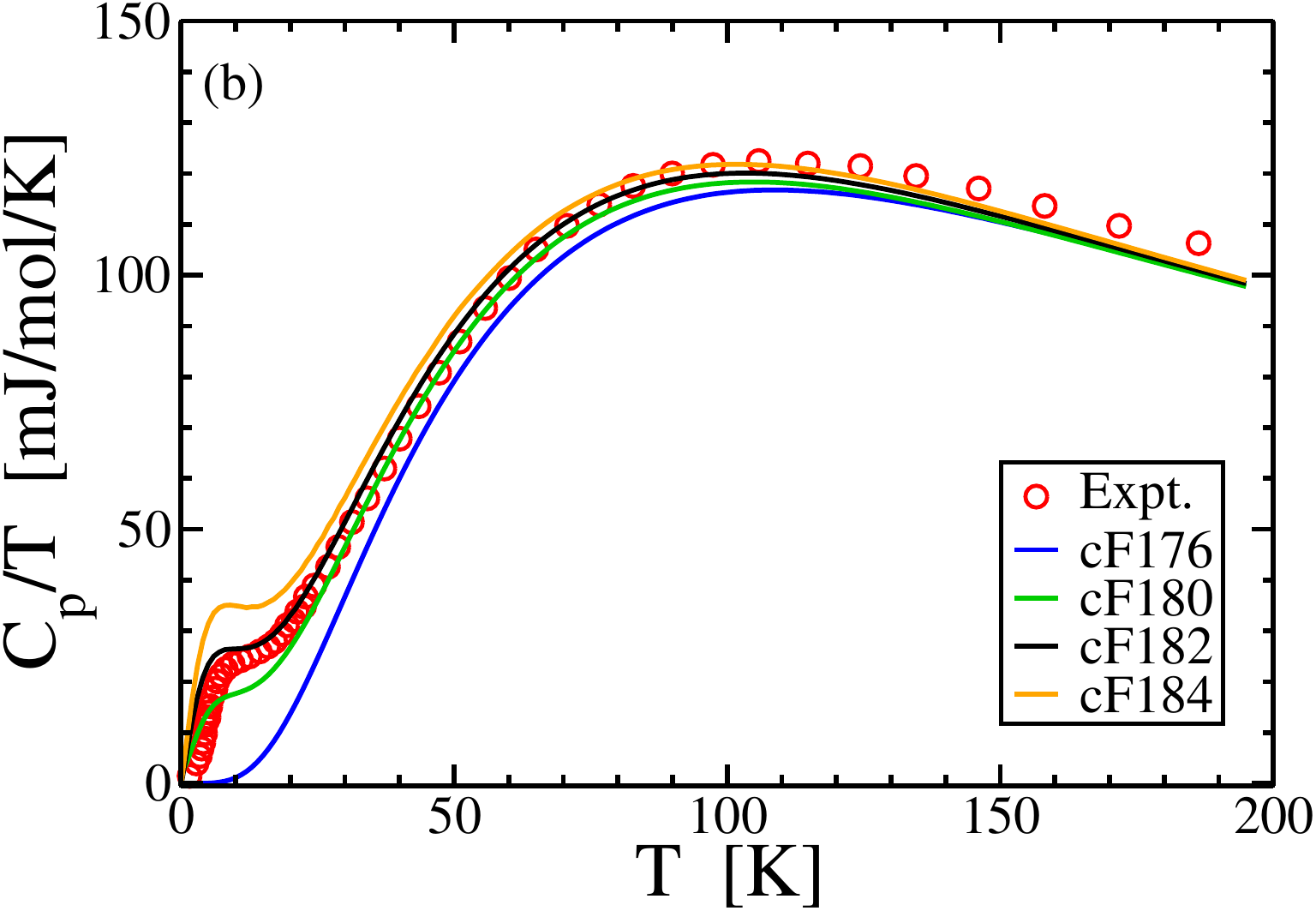}
\caption{(a) Coefficient of linear thermal expansion and (b) specific heat at constant pressure for Al$_{10}$V.cF176 (blue) and Al$_{10}$V.cF180 (green). To approximate Al$_{10}$V structures with 3/4 and full cage filling (Pearson symbols cP182 and cF184), we add 1.5 and 2 times the anharmonic correction to the harmonic phonons of Al$_{10}$V.cF180, shown in black and orange, respectively. Experimental data for Al$_{10}$V with unknown filling are shown as red points~\cite{Safarik2012}.}
\label{fig:LowT}
\end{figure*}

To assess the effectiveness of our method for treating anharmonic guest atom modes, we compare calculated and experimental thermal expansion and specific heat over the temperature range 0--200~K, with particular attention to the low-temperature anomalies. We further isolate the contribution of the guest atoms by contrasting the half-filled cage structure (Al$_{10}$V.cF180) with the empty-cage structure (Al$_{10}$V.cF176).

  Figure~\ref{fig:LowT}(a) shows the linear thermal expansion coefficient $\alpha(T)$ defined in Eq.~\eqref{eq:alpha}, as a function of temperature for Al$_{10}$V.cF176, Al$_{10}$V.cF180, and experimental data for Al$_{10}$V with unknown cage filling. The experimental curve exhibits a pronounced increase in \( \alpha \) in the vicinity of $\Theta_E$. Our anharmonically corrected result for Al$_{10}$V.cF180 captures this anomaly, albeit with somewhat reduced magnitude and lower onset temperature, and shows overall good agreement with the experimental data. In contrast, Al$_{10}$V.cF176 shows no such anomaly, confirming that the anomaly originates from guest atom dynamics.

According to Ref.~\cite{Safarik2012}, each primitive cell contains on average 0.2--0.4 Al atoms occupying the $8a$ site, while Ref.~\cite{Palova2004} estimates a 25\% occupancy. Nevertheless, the precise filling fraction remains uncertain. To model varying filling levels, we scale the anharmonic correction to the harmonic phonon free energy of Al$_{10}$V.cF180. This approximation assumes that the dominant contribution from additional atoms arises through the vibrational term \( F_\mathrm{void} \), and neglects changes to the DFT total energy. The black curve corresponds to 3/4 filling, modeled by adding 1.5 times the anharmonic correction, while the orange curve represents full filling, corresponding to twice the anharmonic correction. As shown in Fig.~\ref{fig:LowT}(a), the black curve provides excellent agreement with experimental data.

Figure~\ref{fig:LowT}(b) shows \( C_p/T \), the specific heat divided by temperature, as a function of temperature, where \( C_p \) is defined in Eq.~\eqref{eq:cp}, for the same systems. Again, the Al$_{10}$V.cF180 result reproduces the low-temperature upturn (below 10~K) observed experimentally, whereas Al$_{10}$V.cF176 shows no such feature. The black (3/4 filled) curve aligns well with the experimental data at low temperatures ($T < 20$~K), while the orange (fully filled) curve offers slightly better agreement at higher temperatures ($T > 20$~K).

Thermal expansion and heat capacity closely track each other in the vicinity of $\Theta_E$. This behavior follows from the thermodynamic identity
\begin{equation}
  \alpha = \frac{\gamma\rho C_V}{K_T}.
\end{equation}
The Gruneisen constant $\gamma$, density $\rho$, and isothermal compressibility $K_T$ vary weakly with temperature; constant volume $C_V$ and constant pressure $C_P$ heat capacities are nearly equal for solids at low temperature. Hence, the anomalous rise in $\alpha$ is approximately proportional to the anomalous rise in $C_P$.

\subsection{Phase diagram}\label{subsection:phase_diagram}

Before presenting our results, we briefly review the experimentally-informed Al–V phase diagrams reported in Refs.~\cite{Murray1989,Gong2004}.
There are two main limitations in such diagrams. First, they are based on a thermodynamic model fit to selected and incomplete experimental data. Second, since thermodynamic equilibrium is difficult to achieve at low temperatures, the lower limits of phase stability are not well established. However, the Al-rich phases are represented as line compounds that are normally assumed to extend down to absolute zero, $T=0$K. As a result, these diagrams often require further confirmation through targeted experimental or computational studies.

Experimentally, the Al-rich phases of Al-V melt peritectically in the vicinity of 940-1010K. We cannot predict melting transitions because we have not modeled the liquid phase. In contrast to the low temperature stability, experimental melting temperatures are likely to be highly accurate.

\begin{figure}[t!]
\centering
\includegraphics[width=\linewidth,clip]{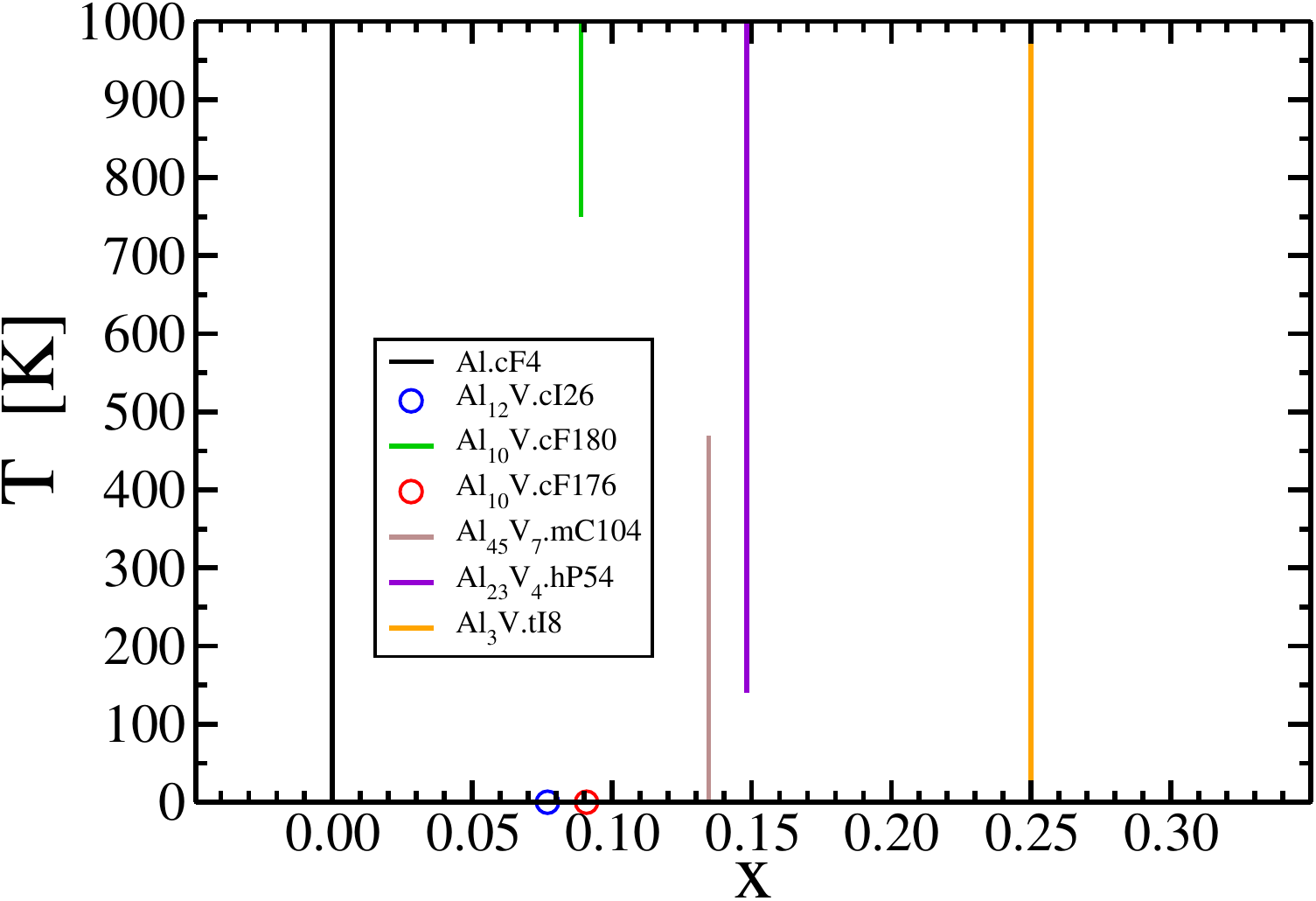}
\caption{(a) Compositional phase diagram of Al–V, showing only phases that compete with Al$_{10}$V. Vertical bars indicate the temperature range over which each phase is predicted to be stable, while circles mark phases that are never stable within the considered range.}
\label{fig:PhaseDiagram}
\end{figure}

Figure~\ref{fig:PhaseDiagram} shows our compositional binary phase diagram of Al–V. As noted earlier, we focus on Al–V phases that compete with Al$_{10}$V. We find that Al$_{10}$V.cF180 is stable for $T > 750$ K but not below. In contrast, Al$_{10}$V.cF176 is never predicted to be stable, indicating that the presence of guest atoms plays a key role in stabilizing this phase.

Regarding other Al–V phases: Al$_{12}$.cI26 is a hypothetical phase in the structure type of Al$_{12}$W, commonly found in the Al-rich part of many Al–transition-metal phase diagrams~\cite{Jahntek2003}. Al$_{23}$V$_4$.hP54 is predicted to be stable for $T > 180$ K. For Al$_{45}$V$_7$.mC104, our calculations show stability in the range $0 < T < 560$ K, somewhat short of the experimental limit of 961 K. However, Al$_{23}$V$_4$.hP54 and Al$_{45}$V$_7$.mC104 lie very close in composition, and their energy differences are small. Resolving their relative stabilities would require calculations with accuracy beyond what is typically achievable with DFT.

\section{Conclusion}\label{conclusion}

We have developed an analytical framework for treating strongly anharmonic vibrational modes where the anharmonicity stems from the motion of guest atoms inside oversized cages. By explicitly modeling the cage potential and solving the associated Schrödinger equation numerically, we incorporated the anharmonic contribution to the vibrational free energy. Either the quantum or the classical free energy may be applied, depending on the temperature range of interest. This approach enabled us to reproduce the anomalous low-temperature behavior of Al$_{10}$V observed experimentally, including the upturns in both the linear thermal expansion coefficient and the specific heat. Furthermore, by incorporating this correction into the free energy, we obtained improved phase stability predictions, showing that the presence of guest atoms is essential for stabilizing the Al$_{10}$V.cF180 structure. 

The system studied here contains only a single anharmonic atom per primitive cell. In systems with multiple such atoms, inter-mode coupling may become significant, introducing additional complexity. Treating these cases would require solving a higher-dimensional Schrödinger equation—formally in $3^N$ dimensions for $N$ anharmonic atoms—or extending the functional form of the fitted potential to capture correlated motion.

Although applied here to a specific compound, the method is broadly applicable to other systems where well-identified, localized anharmonic modes play a central role in thermodynamic behavior. These include clathrates, fullerenes, skutterudites, and other cage-like structures discussed in the introduction~\cite{clathrates,Goto2004,Curl1991,fullerenes,Sales1997,Keppens1998,Borides,Perovskite,Tamura_2002,MOF}. In many of these systems, guest atom vibrations drive unusual and potentially useful physical properties~\cite{BaGeAu,Rogl,AV2Al20,MV2Al20}, making them promising candidates for future application of our approach.

\begin{acknowledgments}
This work was supported by the Department of Energy, USA under Grant No. DE-SC0014506. MM has been supported by Slovak Grant Agency VEGA grant no. 2/0120/25 and by agency APVV grant 23-0281. This research also used the resources of the National Energy Research Scientific Computing Center (NERSC), a US Department of Energy Office of Science User Facility operated under contract number DE-AC02-05CH11231.
\end{acknowledgments}

\appendix

\section{Comparison of DFT Functionals}\label{app:dft_functionals}

\begin{table}[H]
\centering
\caption{Formation enthalpy at $T=0$\,K ($\Delta H_0$) and energy above the convex hull ($\Delta E$) for selected Al–V phases, computed using three different exchange-correlation functionals: PBE, PW91, and HSE06. All energies are in eV/atom and meV/atom, respectively.}
\label{tab:deltaH_functionals}
\begin{tabular}{llcc}
\hline
\textbf{Functional} & \textbf{Phase} & \textbf{$\Delta H_0$} & \textbf{$\Delta E$} \\
\hline
PBE   & Al.cF4            & 0.00 & 0.00 \\
      & Al$_{10}$V.cF176  & -96.17 & 9.56 \\
      & Al$_{10}$V.cF180  & -92.88 & 10.51 \\
      & Al$_{12}$V.cI26  & -72.69 & 16.77 \\
      & Al$_{23}$V$_4$.hP54  & -170.89 & 1.24 \\
      & Al$_{45}$V$_7$.mC104  & -156.56 & 0.00 \\
      & Al$_3$V.tI8 & -292.63 & 0.00 \\
      & V.cI2       & 0.00 & 0.00 \\
\hline
PW91   & Al.cF4            & 0.00 & 0.00 \\
      & Al$_{10}$V.cF176  & -107.10 & 3.24 \\
      & Al$_{10}$V.cF180  & -101.50 & 6.39 \\
      & Al$_{12}$V.cI26  & -81.29 & 12.07 \\
      & Al$_{23}$V$_4$.hP54  & -177.31 & 0.99 \\
      & Al$_{45}$V$_7$.mC104  & -163.39 & 0.00 \\
      & Al$_3$V.tI8 & -282.24 & 0.00 \\
      & V.cI2       & 0.00 & 0.00 \\
\hline
HSE06   & Al.cF4             &  0.00 & 0.00 \\
      & Al$_{10}$V.cF176     & -130.93  & 6.15 \\
      & Al$_{10}$V.cF180     & -124.76  & 9.28 \\
      & Al$_{12}$V.cI26      & -93.93  & 22.06 \\
      & Al$_{23}$V$_4$.hP54  & -217.63  &  5.77  \\
      & Al$_{45}$V$_7$.mC104 & -201.07  &  1.92 \\
      & Al$_3$V.tI8          & -376.99  & 0.00 \\
      & V.cI2                & 0.00  & 0.00 \\
\hline
\end{tabular}
\end{table}

Table~\ref{tab:deltaH_functionals} reports the formation enthalpy at \( T = 0\,\text{K} \), defined as
\begin{equation}\label{eq:enthalpy}
    \Delta H_0 = E_\mathrm{DFT}^{\mathrm{Al}_m \mathrm{V}_n} - \bigl[(1 - x_\text{V}) E_\mathrm{DFT}^{\mathrm{Al}} + x_\text{V} E_\mathrm{DFT}^{\mathrm{V}}\bigr],
\end{equation}
along with the energy above the convex hull, \( \Delta E \), for the Al--V phases considered. Calculations were performed using three exchange-correlation functionals: PBE~\cite{Perdew1996}, PW91\cite{Perdew1993}, and HSE06~\cite{hse06}. While the absolute values of formation energies differ between functionals, the qualitative trends remain consistent: Al$_{10}$V.cF180 lies slightly higher above the hull than Al$_{10}$V.cF176, and both are metastable with respect to decomposition at zero temperature.

Among the three, PW91 predicts the lowest energy above the hull for Al$_{10}$V.cF176 (3.24~meV/atom) and shows the clearest signature of entropic stabilization for Al$_{10}$V.cF180. Unlike PBE, which fails to reproduce the experimentally observed thermal anomalies, PW91 succeeds in capturing the qualitative behavior seen in experiment, motivating our use of this functional throughout the present work. These results underscore the sensitivity of phase stability predictions to the choice of exchange-correlation functional and highlight the limitations of semi-local DFT in accurately capturing subtle energy differences between phases. Given these challenges, a higher-level approach such as the random phase approximation (RPA) may be worth exploring to provide more reliable zero-temperature energetic references.

\section{Quantum vs. classical free energy of the guest atom}\label{app:quantum_vs_calssical_freeEnergy}

We assess the accuracy of the quantum and classical free energy expressions for the guest atom as a function of temperature. For \( T < 250 \,\text{K} \), we use the quantum free energy, ($F_Q$, Eq.~\eqref{eq:void-atom-free-energy}), computed using $n=1000$ eigenvalues and a finite spatial mesh to numerically solve the Schrödinger equation. To correct for the truncation to a finite number of eigenvalues, we fitted the eigenvalues to a power-law function of the form \( a n^{2/3} + b \), where \( a \) and \( b \) are fitting parameters and \( n \) is the eigenvalue index. The exponent \( 2/3 \) is motivated by applying the WKB approximation to a three-dimensional quartic potential. This fit allows us to extrapolate the quantum free energy beyond \( n = 10^4 \) eigenvalues to ensure convergence at 250K. To further correct for finite mesh size, we evaluated the extrapolated free energy at several mesh resolutions and performed a linear extrapolation to the continuum limit (mesh size \( \to 0 \)). We denote the resulting fully converged quantum free energy as \( F_\infty \).

For \( T > 250 \,\text{K} \), we employ the classical free energy, ($F_C$, Eq.~\eqref{eq:classical-free-energy}), which becomes exact in the high-temperature limit. At finite temperatures, the classical free energy can be corrected using the Wigner approximation~\cite{Wigner1932,Deserno2023}, given by
\begin{equation}\label{eq:wigner-correction}
    F_W(T,V) = F_C(T,V) 
    + \frac{\hbar^2}{24 m (k_B T)^2} \left\langle 
    \left( \nabla U(x,y,z) \right)^2 \right\rangle,
\end{equation}
where \( \left\langle (\nabla U)^2 \right\rangle \) is the thermodynamic average of the squared gradient of the potential, with \( U(x,y,z) \) defined in Eq.~\eqref{eq:potential-fit}.

\begin{figure}[t!]
\centering
\includegraphics[width=\linewidth,clip]{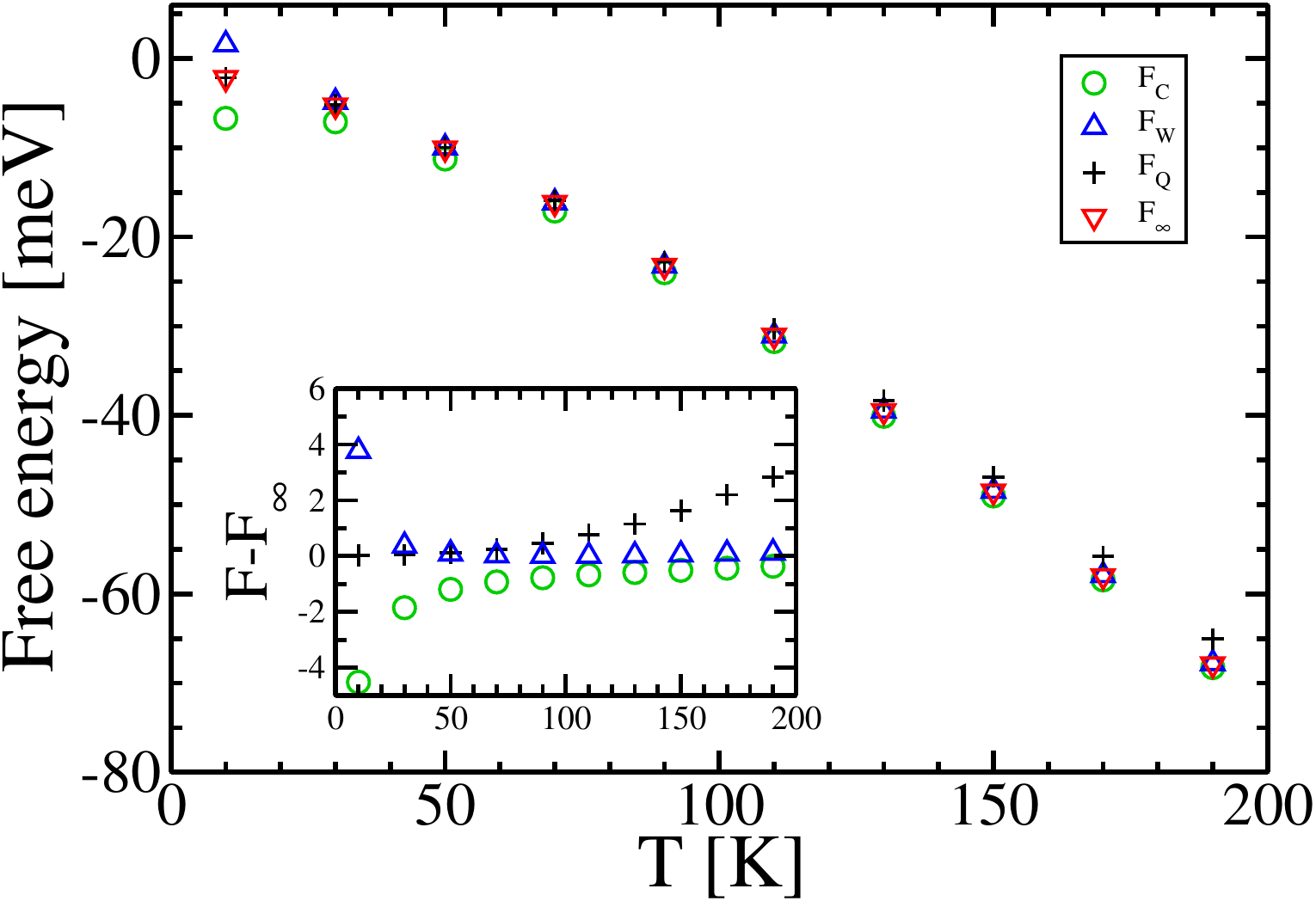}
\caption{Guest atom free energies as functions of temperature, computed using several methods: the classical approximation ($F_C$, Eq.~\eqref{eq:classical-free-energy}, green circles), the classical approximation with Wigner correction ($F_W$, Eq.~\eqref{eq:wigner-correction}, blue triangles), and the quantum expression ($F_Q$, Eq.~\eqref{eq:void-atom-free-energy}, black crosses) evaluated with 1000 eigenvalues and finite mesh size. An extrapolated quantum free energy, denoted \( F_\infty \), obtained by extrapolating the number of eigenvalues to infinity and mesh size to zero, is shown in red triangles. The inset displays the deviation of each method from \( F_\infty \), highlighting the convergence behavior.
}
\label{fig:freeEnergy_convergence}
\end{figure}

Figure~\ref{fig:freeEnergy_convergence} shows the free energy obtained from the above methods for \( T < 200 \,\text{K} \), with the inset displaying deviations from \( F_\infty \), which we take as the exact result. The quantum free energy shows excellent agreement with \( F_\infty \) at low temperatures—where the anomalous behavior of Al$_{10}$V is experimentally observed. However, deviations become noticeable as temperature increases due to the limited number of eigenvalues and finite mesh size, reaching approximately 3~meV at \( T = 200 \,\text{K} \).

In contrast, the classical approximation \( F_C \) and the Wigner-corrected \( F_W \) show initial deviations of approximately $\pm 4$~meV at \( T = 10\,\text{K} \), but improve with increasing temperature. Notably, the Wigner-corrected expression \( F_W \) converges rapidly and reaches near-exact agreement with \( F_\infty \) by \( T = 30\,\text{K} \). Overall, all methods exhibit strong agreement at intermediate temperatures, with residual discrepancies of only a few meV. These results support our hybrid approach of using the quantum free energy for \( T < 250 \,\text{K} \) and the classical approximation for higher temperatures.

\bibliography{Al10V_refs}

\end{document}